\documentclass{article}					

\usepackage{amsmath,amssymb}

\input{diagrams.tex}

\newtheorem{theorem}{Theorem}[section]
\newtheorem{lemma}{Lemma}[section]

\numberwithin{equation}{section}

\newcommand{\AG}{\ensuremath{\mathfrak{A}}}
\newcommand{\FG}{\ensuremath{\mathfrak{F}}}

\newcommand{\NGG}{\ensuremath{\mathfrak{N}}}
\newcommand{\RG}{\ensuremath{\mathfrak{R}}}

\newcommand{\ZG}{\ensuremath{\mathfrak{Z}}}

\newcommand{\DC}{\ensuremath{\mathcal{D}}}

\newcommand{\GC}{\ensuremath{\mathcal{G}}}  
\newcommand{\HC}{\ensuremath{\mathcal{H}}} 
\newcommand{\KC}{\ensuremath{\mathcal{K}}}

\newcommand{\PC}{\ensuremath{\mathcal{P}}}

\newcommand{\VC}{\ensuremath{\mathcal{V}}}

\newcommand{\bp}{\boldsymbol{p}}
\newcommand{\bx}{\boldsymbol{x}}

\newcommand{\gauge}{\text{\textit{$\GC$auge}}}
\newcommand{\obs}{\text{\textit{$\mathcal{O}$bs}}}
\newcommand{\Regione}{\ensuremath{\mathcal{O}}}

\newcommand{\klein}{\mbox{\textrm {\scriptsize Klein}}}
\newcommand{\halmos}{\ensuremath{\raisebox{-.3ex}{\rule{1ex}{1em}}}}

\newcommand{\irr}{\mbox{Irr}}

\begin{document}

\title{\Large\bf Any Compact Group is a Gauge Group}

\author{Sergio Doplicher\thanks{dopliche@mat.uniroma1.it}  
\and  Gherardo Piacitelli\thanks{piacitel@mat.uniroma1.it}}

\date{\footnotesize Dipartimento di Matematica, 
	Universit\`a degli studi di Roma ``La Sapienza'', \\
	P.le A. Moro, 2, 00185 Roma (Italy)}
\maketitle

\centerline{\em Dedicated to Huzihiro Araki on the 
occasion of his $70^{\text{th}}$ birthday.}
\footnotetext{\phantom{k}\\Research supported by MIUR and GNAMPA--INDAM.}
\begin{abstract}
The assignment of the local observables in the vacuum sector, fulfilling the 
standard axioms of local quantum theory, is known to determine uniquely a 
compact group $G$ of gauge transformations of the first kind
together with a central involutive element $k$ of $G$, and a complete 
normal algebra of fields carrying the localizable charges, on which $k$ 
defines the Bose/Fermi grading.

We show here that any such pair $\{G,k\}$, where $G$ is compact metrizable,
does actually appear. The corresponding model can be chosen to fulfill
also the split property.

This is not a dynamical phenomenon: a given $\{G,k\}$ arises as the
gauge group of a model where the local algebras of observables are a
suitable subnet of local algebras of a possibly infinite product of
free field theories. \\[2em]

\end{abstract}

\section{Introduction}

When few standard assumptions of Local Quantum Physics on the physical, 
four dimensional Minkowski 
spacetime
are fulfilled, the assignment of the local
observables is sufficient to construct an algebra \FG\ of fields
carrying localizable charges and a compact group $G$ of gauge transformations of the first 
kind, where $\FG$ is complete
in the sense that {\em all} superselection
sectors with finite statistics (intrinsically determined by the local 
observables) can be reached acting with the fields
on the vacuum~\cite{doplicher_90b}. 
Moreover, the construction is unique 
if in addition one requires normal Bose/Fermi commutation relations 
at spacelike separations; the corresponding
$\mathbb{Z}_2$--grading is  defined
by a central involution $k\in G$. The local observables are then identified 
with the gauge invariant part of $\FG$, and the superselection sectors are in 
one to one correspondence with the classes of irreducible continuous unitary 
representations of the gauge group $G$. 
More
precisely, $\FG$ is constructed from a cross product of $\AG$ by the
superselection category,
and $G$ appears as the group of all automorphisms of $\FG$ leaving $\AG$ 
pointwise
fixed. Representations of $\AG$ obeying the selection criterion of
localizability and with finite statistics are described by localized
morphisms of $\AG$ induced by finite dimensional $G$ invariant Hilbert 
spaces in
$\FG$ with support $I$; the correspondence between a localized morphism and the
representation of $G$ obtained as the restriction of  gauge
automorphisms to the corresponding Hilbert space in $\FG$ provides a natural
isomorphism between the superselection category and a category of
representations of $G$, equivalent to the category of all finite
dimensional unitary continuous representations. For a survey see 
e.g.~\cite{kyoto,capri}.

The proof of this reconstruction theorem was closely related to a new
duality theory for compact groups~\cite{doplicher_89b}.
The gauge group $G$  will be metrizable\footnote{Equivalently, the compact 
group $G$ has a countable basis of open sets, or has a countable set of
equivalence classes of irreducible continuous unitary 
representations.}
precisely in theories with at most
countably many superselection sectors, hence, 
in particular, in theories where the 
split property is fulfilled by the field net~\cite{doplicher_84}.
So it is quite natural to ask which compact groups may arise as gauge groups.
Indeed, here we provide a natural, functorial construction which maps
any given system $\{G,k,\mu\}$~---~with $G$ a metrizable compact group,
$k\in\ZG(G)$, $k^2=e$, and $\mu$ a suitable mass function over a
generating subset of the spectrum of $G$, cf below~---~to 
an observable net with a distinguished vacuum sector,
admitting precisely $G$ as its gauge group, and $k$ as the grading element. 
Moreover, the field net will fulfill the split property. 
 
Let $\irr(G)$ be the set of all irreducible subrepresentations of the
regular representation.  
We say that a subset
$\Delta\subset\irr(G)$ is symmetric if it is stable under 
conjugation, generating if every element of $\irr(G)$ appears as
a subrepresentation of tensor products of elements of $\Delta$
\phantom{k}\footnote{By the Stone--Weierstrass theorem, a symmetric 
$\Delta$ is generating if and only if it is faithful, i.e.~it separates points
of $G$.}. 
In what follows, by a mass function we shall always
understand a map  $\mu:\Delta^{(\mu)}\longrightarrow
(0,\infty)$ with $\Delta^{(\mu)}\subset\irr(G)$ a symmetric,
generating subset associated to mutually orthogonal subspaces, and such 
that $\inf\{\mu(\Delta^{(\mu)})\}>0$; moreover, 
$\mu$ will satisfy the growth condition
\begin{equation}
\label{cond}
\forall \delta>0,\;
\sum_{\sigma\in\Delta^{(\mu)}}d(\sigma)\mu(\sigma)^{4}\;e^{-\mu(\sigma)\delta}
<\infty
\end{equation}
($d(\sigma)=\mbox{dim}(K_{\sigma})$ is the dimension of the
repre\-senta\-tion~$\sigma\in\Delta^{(\mu)}$ acting on the Hilbert space 
$K_\sigma\subset L^2(G)$), 
which will imply the split property for the field net of the 
associated model.

A gauge triple is a system $T=\{G,k,\mu\}$ where $G$ is a compact, metrizable
group, $k\in G$ an involutive central element and $\mu$ a mass function. 
A continuous, surjective group homomorphism  $G\rTo^{\eta}G_1$ 
defines an arrow
$$
\{G,k,\mu\}\rTo^{\eta}\{G_1,k_1,\mu_1\}
$$
in the category \gauge\ of gauge triples if and only if
$$
\eta(k)=k_1
$$
and 
the transposed map $\irr(G_1)\rTo^{\eta_*}\irr(G)$, 
$\eta_*\sigma=\sigma\circ\eta$, fulfills the conditions
\begin{eqnarray*}
\eta_*\Delta^{(\mu_1)}&\subset&\Delta^{(\mu)},\\
\mu\circ\eta_*&=&\mu_1.
\end{eqnarray*}

We shall construct a functor $F$ from $\gauge$ to the category $\obs$\
of covariant observable (i.e. local) nets in a vacuum 
sector. The objects of $\obs$ are the systems
$\{\HC,\AG,U\}$, where $\AG$ is an irreducible, additive local 
net of von Neumann 
algebras on the separable Hilbert space $\HC$, defined on the set
$\KC$ of double cones in Minkowski space, and fulfilling duality\footnote{We
recall that the net  
$\AG$ is said to fulfill duality if, for any double cone $\mathcal O$,
$\AG(\Regione)=\AG(\Regione')'$.}; $U$ is a 
strongly continuous unitary representation of the Poincar\'e group 
$\mathcal P$, covariantly acting on $\AG$, 
satisfying the spectrum condition and having an
invariant vector (the vacuum vector), unique up to a phase. An arrow 
$$
\{\HC_1,\AG_1,U_1\}\rTo\{\HC,\AG,U\}
$$
is given by an isometry $V:\HC_1\rightarrow\HC$ and an inclusion of nets
$\Phi:\AG_1\hookrightarrow\AG$ such that
\begin{gather*}
VA=\Phi(A)V,\quad A\in\AG_1,\\
VU_1(L)=U(L)V,\quad L\in\PC.
\end{gather*}
Hence, $\AG_1\subset\AG$ in the sense  of~\cite{conti_00}.

The results of~\cite{doplicher_90b}\ provide us with a reconstruction functor 
$r$ which
associates to any system $\{\HC,\AG,U\}$ in $\obs$ 
the corresponding couple $\{G,k\}$
where $G$ is the gauge group and
$k$ is the element grading the commutation rules.

We are ready to state our main result.
\begin{theorem}
\label{funtore}
There exists a faithful, contravariant functor 
$$
\gauge\rTo^F\obs
$$
such that the following diagram is commutative
\begin{diagram}
\{G,k,\mu\}	&		&\rTo^F	&		&\{\HC,\AG,U\}	\\
		&\rdTo_{f}	&	&\ldTo>{r}	&		\\
		&		&\{G,k\}&		&		\\
\end{diagram}
where $f$ forgets the mass
function. Moreover, for any gauge triple $T$, the canonical field net
associated with $F(T)$ fulfills the split property.

\end{theorem}
We recall here that the split property selects the models which are more
relevant for physics among theories fulfilling the general axioms of 
locality, covariance, and spectrum condition; in particular it
allows us to derive rigorously a weak form of the Noether theorem 
and an exact variant of the current algebra
\cite{doplicher_82,doplicher_83,buchholz_86}; this provides a natural
approach towards a full Quantum Noether Theorem (see~\cite{carpi_99}\
for partial results in conformal models). 

{\em Remark 1.1.} 
Let $T=\{G,k,\mu\}$ be a gauge triple, and $F(T)=\{\HC,\AG,U\}$;
$\AG$ will be obtained as the $G$-invariant part of a field net $\FG$
associated to $T$. 
$\FG$ will be generated , as $\sigma$ runs through $\Delta^{(\mu)}$,
by $d(\sigma)$ independent free fields of mass $\mu(\sigma)$, acting on a 
Hilbert space $\mathcal H_\sigma$, of scalar
or Dirac type according to whether $\sigma(k)=+1$ or $-1$ respectively,
and fulfilling normal commutation relations; thus fields associated to
$\sigma\neq \sigma'$ anticommute if $\sigma(k)=\sigma'(k)=-1$, and commute 
otherwise. The net $\FG$ will act irreducibly over the (infinite, if 
$\Delta^{(\mu)}$ is infinite) tensor product (relative to the sequence of 
vacuum state vectors) over $\sigma\in\Delta^{(\mu)}$ of the Hilbert spaces 
$\mathcal H_\sigma$.
To prove that $G$ is indeed the full gauge group associated to 
$\AG$, it suffices to prove that the field net $\FG$  fulfills 
the split property, together with a further cohomological 
condition~\cite[Theorems 4 and 2, \S 3.4.5.]{roberts_90}. 
Actually, one might conjecture that the 
infinite tensor product of full free theories has no nontrivial 
sectors\footnote{A related conjecture has been formulated by 
Roberto Longo~\cite{longo_conjecture}; we defer this problem to future 
investigation.}. Were this conjecture  
true, the extension of the functor $F$ to a larger category of gauge 
triples, defined by dismissing the growth condition on the mass functions,
still would make the above diagram commutative, without any need to invoke and
establish the split property of the infinite tensor product model. Furthermore,
this might allow us to extend the present result to theories with uncountably
many superselection sectors, where the field algebra cannot fulfill the 
split property, but which however might hardly have any physical interest.

{\em Remark 1.2.} Let $T,T_1$ be two gauge triples, and $\FG,\FG_1$ the canonical field nets 
associated with $F(T),F(T_1)$, respectively. If there is an arrow 
$T\rTo^{\eta}T_1$, then $F(T_1)$ is a subsystem of $F(T)$ 
and one has the following 
commuting square of inclusions of nets (in the sense of~\cite{conti_00})
$$
\begin{array}{ccc}
\AG_1&\subset&\FG_1\\
\cap&&\cap\\
\AG&\subset&\FG
\end{array}
$$
Then 
\begin{align*}
\FG_1\vee\AG&=\FG^{\ker\eta},\\
\FG_1\wedge\AG&=\FG_1^{G_1},
\end{align*}
a statement about the superselection theory for subsystems, 
which holds true in general. See~\cite{conti_00}.

The construction of our models is outlined in section \ref{costr_F}, 
and the proof of the main theorem 
is completed in section \ref{split}.

\section{Construction of the models}
\label{costr_F}

A mass function $\mu$ 
naturally defines a faithful representation $D$ of $G$ as the 
direct sum of the representations $\sigma$ contained in its domain 
$\Delta^{(\mu)}$, with representation space $K\subset L^2(G)$ stable 
under the canonical conjugation on $L^2(G)$, which induces a conjugation 
$J$ of $K$.
The r\^ole of the grading element $k$ 
is to select Fermi and Bose sectors: since $k$ is central
and involutive, we can decompose 
$\Delta^{(\mu)}=\Delta_+^{(\mu)}\cup\Delta_-^{(\mu)}$, 
where  $\sigma(k)=\pm I_{K_\sigma}$, $\sigma\in\Delta^{(\mu)}_{\pm}$.
Correspondingly, the space $K$ has the natural decomposition 
$K=K_+\oplus K_-$ as the direct sum  of the two eigenspaces of the 
selfadjoint, unitary operator 
$D(k)$.

We consider two generalized free fields, a scalar field $\phi$ defined
on the $K_+$-valued test functions, and a Dirac field $\psi$ 
defined on
the $K_-\otimes\mathbb C^4$-valued  test functions (a brief reminder is 
included in the appendix, for convenience of the reader). 
Define $\FG$ as the tensor product of the nets
$\FG_-$ and $\FG_+$ associated to $\phi$ and $\psi$ respectively; 
the action of
$G$ on the fields is induced by the natural pointwise actions of $G$\
on the test functions, and the mass function determines the energy--momentum 
spectrum. 
The observable net $\AG=\FG^{G}$ is the net of
gauge invariant elements. The vacuum representation of \AG\ is
generated by the restriction to \AG\ of the vacuum state of \FG.  

The field algebra \FG\ is thus
a product of massive free field models, on which the elements of
the desired group $G$ act as gauge transformations on multiplets of
fields and $k$ determines
the Bose/Fermi grading. For each $\sigma\in\Delta^{(\mu)}$,
there is a multiplet of $d(\sigma)$ fields on which $G$ acts and defines
an unitary representation equivalent to $\sigma$. Since the vacuum is cyclic
and $G$-invariant, 
the action of $G$ on the fields is implemented by a faithful, 
strongly continuous 
unitary representation \VC\ of $G$. The commutation relations are 
graded by $\VC(k)$. The basic properties of locality, covariance, and 
spectrum condition~\cite{haag,araki} are evidently fulfilled, only the proof
of the duality property of the net $\mathfrak A$ in its vacuum sector has to 
be sketched. 

We first prove twisted duality for the field net
$\FG$. The net of fixed points under the action of a gauge 
group $G_{\text{max}}$, {\em a priori} larger than
$G$, will be the tensor product of nets, each fulfilling duality,
hence it will fulfill duality as well 
(cf e.g.~\cite [Lemma 10.1]{doplicher_84}). 
This implies (\cite[Theorem 3.6]{doplicher_90b})
that \FG\ fulfills twisted duality, which in turn implies that for any 
closed subgroup $G\subset G_{\text{max}}=U(\HC)\cap\FG'$, the fixed point net
$\AG=\FG^G$ fulfills duality in the vacuum~\cite{doplicher_69}. 

To be more precise, the Tychonov product $G_{\text{max}}$ of the full unitary 
groups $U(K_\sigma)$
(or of the full orthogonal group $O(K_\sigma)$ if $\sigma=\overline{\sigma}$)
is represented on $K$ by the diagonal action in the decomposition
of $K$ as the direct sum of the subspaces
$K_\sigma$; composing this representation
with the second quantization functor $\Gamma$ (which intertwines direct sums 
and tensor products, cf remark 1 above) provides a unitary 
representation of $G_{\text{max}}$ which induces gauge automorphisms on $\FG$,
generated by the products of the actions of the full gauge groups $U(K_\sigma)$
(or $O(K_\sigma)$). 

The subnet $\AG_{\text{min}}=\FG^{G_{\text{max}}}$\
of fixed points under $G_{\text{max}}$  is now the tensor product of the 
nets $\AG_{\sigma}:=\FG_{\sigma}^{U(K_{\sigma})}$ (or 
$\FG_{\sigma}^{O(K_{\sigma})}$) of
fixed points in the  field algebra generated by the free fields
$\phi_{\sigma}$ or $\psi_\sigma$.  Since each
$\AG_{\sigma}$ fulfills duality in its vacuum sector, so
does the tensor  
product net $\AG_{\text{min}}$, hence $\AG$ too.

We thus defined an object of $\obs$ associated by $F$ to an object
of $\gauge$; to an arrow $\{G,k,\mu\}\rTo^{\eta}\{G_1,k_1,\mu_1\}$
in $\gauge$, we associate an arrow $F(\eta)=(V,\Phi)$ in $\obs$ as follows.
Let $v_\eta$ be the restriction to $K_1$ of the isometry from $L^2(G_1)$
to $L^2(G)$ given by the transposition of $\eta$, and define the 
isometry $\widetilde V$ and the *-homomorphism 
$\widetilde\Phi:\FG_1\longrightarrow\FG$ by
\begin{gather*}
\widetilde V:=\Gamma(v_\eta),\quad
\widetilde \Phi\big(e^{i\phi(f)}\big):=
e^{i\phi(v_\eta f)},\quad
\widetilde \Phi(\psi(f)):=\psi(v_\eta f),
\end{gather*}
where $\phi, \psi$ is the generating Bose, resp.~Fermi 
generalized free field, for all
the appropriate test functions $f$. The desired arrow $F(\eta)$
is obtained as the restriction of $\widetilde V,\widetilde\Phi$ respectively
to the gauge invariant subspace and subalgebra. A routine check shows that
$F(\eta_1\circ\eta_2)=F(\eta_2)\circ F(\eta_1)$ whenever 
$\eta_1\circ\eta_2$ is defined, and $F$ is indeed a contravariant functor. 
By Theorem 2.2 of \cite{conti_00} and the comments appended there, $F(\eta)$
in turn determines $\eta$, and $F$ is faithful.

\section{Split property and completeness of the superselection theory}
\label{split}

In this section we show that the net $\FG$ constructed as in the appendix
from  the gauge triple $\{G,k,\mu\}$ is the canonical complete field net
associated to $\FG^G$. To this end, we have to show that it fulfills the 
split property and the cohomological condition~(\ref{aggettivo}) below
(\cite{roberts_90}).

The arguments for the split property 
are adaptations from the discussion
of the Bose case~\cite{driessler_79,frohlich_76,d'antoni_83,d'antoni_87}. 
In particular we prove that the split property holds if the mass function
fulfills the growth condition (\ref{cond}).

We recall that a net $\RG$ is said to fulfill the split property
if, for any inclusion of double
cones $\Regione_1\subset\Regione_2$ such that  
the closure of
$\Regione_1$ is contained in the open set $\Regione_2$ (in symbols: 
$\Regione_1\subset\subset\Regione_2$), then the  inclusion
$\RG(\Regione_1)\subset\RG(\Regione_2)$ is split, that is, there exists a
type I factor \NGG\
such that $\RG(\Regione_1)\subset\NGG\subset\RG(\Regione_2)$. If the canonical
field net associated with an observable net fulfills the split
property, then the observable net also fulfills the split property,
but sufficiently general conditions that imply the converse are not 
known~\cite{doplicher_82}.

The field net $\FG$ is the tensor product 
of the nets $\FG^{(\pm)}$; since finite tensor products 
preserve the split property, it is enough to show that $\FG^{(\pm)}$ both
satisfy the split property. In the case of $\FG^{(+)}$, 
condition \ref{cond}\ (restricted to $\Delta_+^{(\mu)}$)
is sufficient by the results of~\cite{doplicher_84,d'antoni_87}.
Here we show that the same condition (restricted to $\Delta_-^{\mu}$)
is sufficient also in order to ensure the split property for $\FG^{(-)}$.

As a first step, we give an alternative proof of the split property for
the field net generated by a massive Dirac free field, which was first proved
in~\cite{summers_82}. The 
argument is an adaptation of the one given in~\cite{d'antoni_83}\ for
the Bose case, and 
provides an estimate analogous to the one given in~\cite{d'antoni_87}; for the
reader's convenience, we give some details and point out the main differences.

Let us consider  
the doubled theory of the massive Dirac free field (followed
by a Klein transformation); this theory is induced by the fields 
$$
\psi_{1}=\psi_{m}\otimes I,\hspace{4ex} \psi_{2}=V\otimes\psi_{m},
$$
where $Q$ is the usual charge and $V=e^{i\pi Q}$. 
We now choose as gauge transformations
\begin{eqnarray*}
\alpha_{\theta}:\psi_{1}&\mapsto&\cos(\theta)\psi_{1}+\sin(\theta)\psi_{2}\\
\alpha_{\theta}:\psi_{2}&\mapsto&-\sin(\theta)\psi_{1}+\cos(\theta)\psi_{2},
\end{eqnarray*}
for which the conventional Noether theorem gives the conserved current
$$
j^{\mu}=\frac{i}{2}\{\psi_{1}\gamma^{\mu}\psi_{2}-
		\psi_{2}\gamma^{\mu}\psi_{1}\}.
$$
Equivalently, with $\eta=(\eta_1,\eta_2)$, we can write
$\psi(f,\eta)=\eta_1\psi_1(f)+\eta_2\psi_2(f)$, so that
$\alpha_\theta(\psi(f,\eta))=\psi(f,R(\theta)\eta)$, with $R(\theta)$
the orthogonal rotation by $\theta$.
 
Note that $\alpha_{\pi/2}(F\otimes I)=(I\otimes F)^{\klein}=
I\otimes F_{+}+V\otimes F_{-}$ where $F=F_{+}+F_{-}$ is the 
decomposition of $F$ into the sum of a bosonic and a fermionic
operator, $F_{\pm}=(F\pm VFV)/2$.  
Let $D_{r}=\{(0,\bx)\in\mathbb{R}^{4}:|\bx|<r\}$ and 
$\Regione_{r}={D_{r}}''$ be the double cone of radius $r$, centred at the 
origin; then $\Regione_{r}\subset\subset
\Regione_{r+\delta}$, $\delta>0$.
Thanks to covariance, it suffices to show that the inclusion
$\FG(\Regione_{r})\subset\FG(\Regione_{r+\delta})$ is split, $r,\delta >0$. 
Standard techniques (see 
\cite{d'antoni_83}\ for the Bose case) allow one to give sense to
$j^{\mu}(f)=\int d^{4}x\; j^{\mu}(x)f(x)$ as a selfadjoint 
operator for real test functions $f$, and show that, for a suitable $f$,
$e^{i\theta j^{0}(f)}$ is in 
$\FG(\Regione_{r+\delta})$ and implements 
$\alpha_{\theta}$ on $\FG(\Regione_{r})$. The $\mathcal C^{\infty}$ functions 
$f$ will have the 
form $f(x^{0},\bx)=h(x^{0})g(\bx)$, with
$g$ supported in $\{\bx:|\bx|<r+2\frac{\delta}{3}\}$, $g(\bx)=1$ in 
$D_{r+\frac{\delta}{3}}$,
and $h$ with support in $[-1,1]$ 
and such that $\int h(s)ds=1$; note that the desired action
of $j^{0}(f)$\
does not depend on the smearing in time, i.e. on the choice of $h$\
\cite[Lemma I]{kastler_66}. In what follows, we shall write 
$$
J_{m}=\frac{\pi}{2}j^{0}(f).
$$

Here the argument differs from~\cite{d'antoni_83}, due to the presence
 of the Klein transformation ($\alpha_{\pi/2}$\
is not the flip any more).
Consider the algebraic *-isomorphism
\begin{displaymath}
\sigma\left(\sum_{i=1}^{n}F_{i}F'_{i}\right)=\sum_{i=1}^{n}F_{i}\otimes F'_{i}
\end{displaymath}
between the *-algebra generated by 
$\FG_{m}(\Regione_{r})\FG_{m}(\Regione_{r+\delta})'$ and the algebraic
tensor product
$\FG_{m}(\Regione_{r})\odot\FG_{m}(\Regione_{r+\delta})'$.
We wish to find two normal, faithful states 
$\lambda$ and $\ell$, such that
\begin{equation}
\label{iso}
\lambda\left(\sum_{i=1}^{n}F_{i}F'_{i}\right)=
	\ell\left(\sum_{i=1}^{n}F_{i}\otimes F'_{i}\right),
\end{equation}
where the $F_{i}$'s  (resp.~$F_{i}'$'s) are arbitrarily chosen in 
$\FG_{m}(\Regione_{r})$ (resp.~$\FG_{m}(\Regione_{r+\delta})'$).
This would imply that $\sigma$ could be extended  to a unique
normal *-isomorphism 
$\bar\sigma:\FG_{m}(\Regione_{r})\vee\FG_{m}(\Regione_{r+\delta})'\rightarrow
\FG_{m}(\Regione_{r})\overline\otimes\FG_{m}(\Regione_{r+\delta})'$, which 
implies the split property. 
Taking the vector 
$\Phi=e^{-{i} J_{m}}\;\Omega\otimes\Omega$, we define on
$\FG_{m}(\Regione_{r})\vee\FG_{m}(\Regione_{r+\delta})'$ the normal state
$\lambda(F)=(\Phi,F\otimes I\;\Phi)$. Then, we define the normal 
positive map $X$ 
of $\FG_{m}(\Regione)\overline\otimes\FG_{m}(\hat\Regione)'$ into itself,
extending
$X(T\otimes S)=T_-\otimes S_-$ by linearity and continuity.
Finally we define
$\Psi=e^{-2{i} J_{m}}\;\Omega\otimes\Omega$, and the linear
functional
\begin{displaymath}
\ell(\cdot)=(\Omega\otimes\Omega,\cdot\;\Omega\otimes\Omega)+
	(\Psi,X(\cdot)\Omega\otimes\Omega),
\end{displaymath}
on $\FG_{m}(\Regione_{r})\overline\otimes\FG_{m}(\Regione_{r+\delta})'$.
The functionals $\lambda$ and $\ell$ are ultraweakly continuous by 
construction;
$\lambda$ is evidently positive, normalized, and faithful. Also
$\ell$ will be  normalized, positive and faithful if~(\ref{iso}) holds.
If $F\in\FG_{m}(\Regione_{r})$ and 
$F'\in\FG_{m}(\Regione_{r+\delta})'$, we denote
$F=F_{+}+F_{-},\ F'=F'_{+}+F'_{-}$ where  
$F_{\pm}=(F\pm VFV)/2$ and analogously for $F'$; then 
$\lambda(FF')=\lambda(F_{+}F'_{+})+\lambda(F_{+}F'_{-})+
\lambda(F_{-}F'_{+})+\lambda(F_{-}F'_{-})$. Since $F'\otimes I$ commutes with
$e^{i J_{m}}$ and Fermi fields have zero vacuum expectation values, we have
$\lambda(F_{+}F'_{+})=\omega_{0}(F) \omega_{0}(F')$ and $\lambda(F_{-}F'_{+})=0$;
since $e^{{i} J_{m}}$ commutes with $U_{\pi}$\
(the unitary operator implementing $\alpha_{\pi}$), we have
$\lambda(F_{+}F'_{-})=(U_{\pi}\Omega\otimes\Omega,e^{{i} J_{m}}
(F_{+}F'_{-}\otimes I)e^{-{i} J_{m}}U_{\pi}\Omega\otimes\Omega)=
-\lambda(F_{+}F'_{-})=0$.
Thus, 
\begin{gather*}
\lambda(FF')=\omega_{0}(F)\omega_{0}(F')+\lambda(F_{-}F'_{-})=\\
=
\omega_{0}(F)\omega_{0}(F')+(\Omega\otimes\Omega,e^{{i} J_{m}}(F'_{-}\otimes I)
e^{-{i} J_{m}}(I\otimes F_{-})\Omega\otimes\Omega),
\end{gather*} 
where we used that
$e^{{i} J_{m}}(F_{-}F'_{-}\otimes I)e^{-{i} J_{m}}=
e^{{i} J_{m}}(F_{-}\otimes I)e^{-{i} J_{m}}(V\otimes F'_{-})$,
and $V\Omega=\Omega$. Since $(F'_-\otimes I)e^{iJ_m}=e^{-iJ_m}(F'_-\otimes I)$, 
we have 
$\lambda(F_-F'_-)=(\Psi,(F'_{-}\otimes F_{-})\;\Omega\otimes\Omega)$, hence
equation~(\ref{iso}). 

The former was an alternative proof of the split property
for the Fermi--Dirac model; 
we are now interested in getting the estimate
\begin{equation}
\label{stima}
\|J_{m}(f)\Omega\otimes\Omega\|^{2}\leqslant
\mbox{const}\left(r+\frac{\delta}{2}\right)^{6} 
\left(\frac{m}{\delta}\right)^{3}
\left(\frac{m\delta}{2}+2\right)^{4}e^{-m\delta},
\end{equation}
where we recall the r\^ole of $r$ and $\delta$, 
$\Regione_{r}\subset\subset\Regione_{r+\delta}$. 
To this purpose we can slightly modify
the estimate of 
\cite[Lemma C.1]{d'antoni_87}\ to obtain, for all positive $a$'s:
\begin{equation}
\label{stima_2}
\inf_{\begin{array}{c}
{\scriptstyle h\in\mathcal{D}(-1,1)}\\{\scriptstyle \int h=1}
\end{array}}
\sup_{\omega\geqslant a}|\omega^{2}\hat h(\omega)|\leqslant
4a^{\frac{1}{2}}\left(\frac{a}{2}+2\right)^{2}e^{-\frac{a}{2}}.
\end{equation}
Then standard calculations give
\begin{displaymath}
\|J_{m}\Omega\otimes\Omega\|^{2}\leqslant
\mbox{const}\int_{4m^{2}}^{\infty}d\kappa^{2}
\left(1-\frac{4m^{2}}{\kappa^{2}}\right)^{\frac{1}{2}}
\int d^{3}\bp\sqrt{|\bp|^{2}+\kappa^{2}}\left|
\hat f(\sqrt{|\bp|^{2}+\kappa^{2}},\bp)\right|^{2},
\end{displaymath}
where, as usual, 
$\hat f(p)=\frac{1}{(2\pi)^2}\int d^{4}x\;f(x)e^{ix_{\mu}p^{\mu}}$.
Now, proceeding as in~\cite{d'antoni_87}, we obtain
\begin{displaymath}
\|J_{m}\Omega\otimes\Omega\|^{2}\leqslant
\mbox{const}\;m^{2}{2}\left(r+\frac{\delta}{2}\right)^{6}
\left(\frac{2}{\delta}\right)^{4}
\sup_{\omega\geqslant m\delta}|\omega^{2}\hat{h}(\omega)|^{2},
\end{displaymath}
and using~(\ref{stima_2})\ and the fact that $J_{m}$ does not depend
on the choice of $h$, we obtain the desired 
estimate~(\ref{stima}).
 
We are now ready to prove the split property for $\FG^{(-)}$.
The doubled theory $(\FG^{(-)}\otimes\FG^{(-)})^{\klein}$ is generated by the
Dirac fields $\psi_{\sigma,i}(f,\eta)$ of mass $\mu(\sigma)$\
(where $\sigma\in\Delta_-^{(\mu)}$\
and $i=1,2,\dots,\text{dim}(K_{\sigma})$),
with the addition of the internal degrees of freedom
$\eta\in\mathbb{C}^2$. Defining the gauge transformation 
$$
\alpha_\theta\left(\psi_{\sigma,i}(f,\eta)\right)=
\psi_{\sigma,i}(f,R(\theta)\eta),
$$
where $R(\theta)$ is the rotation by $\theta$, we find as generator of
such rotations the operator $J=\sum_{\sigma\in\Delta_-^{(\mu)}}\sum_{i=1}^{\text{dim}K_{\sigma}}J_{\sigma,i}$ where the 
$J_{\sigma,i}$'s strongly commute with each other.
The only remaining question is whether $J$  
exists in the incomplete infinite tensor product. Writing $J_m$ for the local
generator of the rotation of a doubled Dirac field of mass $m$, 
this is the case
if 
$\sum_{\sigma\in\Delta^{\mu}}
d(\sigma)\|J_{\mu(\sigma)}\Omega\otimes\Omega\|
<\infty$, as can be seen applying the following lemma.
Using estimate~(\ref{stima}) we get the  desired condition~(\ref{cond}), restricted to $\Delta_-^{(\mu)}$.
\begin{lemma}
\label{inf_gen}
Let $\{J_{n}, n\in\mathbb{N}\}$, be a family of strongly commuting selfadjoint 
operators affiliated to the von Neumann algebra \RG\ on the Hilbert space
\HC; if
$\Omega\in\bigcap_n\DC(J_{n})$ is a cyclic vector for $\RG'$ and 
if $\sum\|J_{n}\Omega\|<\infty$, then $U(\lambda):=\prod_{n=1}^{\infty}
e^{{i} J_{n}\lambda}$ defines a strongly continuous one parameter group 
with generator $J=\sum_{n}J_{n}$.
\end{lemma}
{\em Proof}.
For any real $\lambda$ and $T\in\RG'$, the sequence
\begin{displaymath}
x_{n}(\lambda)=\left(\prod_{k=1}^{n}e^{{i}\lambda J_{k}}\right)T\Omega=
T\left(\prod_{k=1}^{n}e^{{i}\lambda J_{k}}\right)\Omega
\end{displaymath}
fulfills 
$\|x_n(\lambda)-x_m(\lambda)\|\leqslant\|T\||\lambda|\sum_{k=m+1}^n\|J_k\Omega\|$,
where we used repeatedly the unitarity and the triangle inequality,
and finally that $|e^{it}-1|\leqslant|t|$ together with the functional calculus.
Hence, there exists
$x(\lambda)=\lim_{n\rightarrow\infty}x_{n}(\lambda)$,
and $\|x(\lambda)-x_{n}(\lambda)\|\leqslant\| T\| |\lambda|\sum_{k=n+1}^{\infty}
\|J_{k}\Omega\|$; a ``$3\varepsilon$~argument'' gives
continuity in norm of $\lambda\mapsto x(\lambda)$.
Since $\RG'\Omega$ dense in \HC, 
$U(\lambda)=\mbox{s-}\!\lim_{n\rightarrow\infty}
\prod_{k=1}^{n}e^{{i}\lambda J_{k}}$ exists for any $\lambda$ and
$\lambda\mapsto U(\lambda)$ is strongly continuous.  ~\halmos\\*[2ex]

By the results of~\cite{roberts_90}, a field net $\FG$ fulfilling the split 
property is a complete field net for the fixpoint net $\FG^G$
if it also fulfills the following condition. Let $p$ be a path 
connecting the double cones $\Regione_{0}$ and $\Regione$, i.e.
 a collection of $2n+1$  double cones 
$\Regione_{0},\Regione_{1},
\cdots\Regione_{n}=\Regione$, $\hat\Regione_{1},\cdots,\hat\Regione_{n}$\
such that $(\Regione_{i-1}\cup\Regione_{i})\subset\hat\Regione_{i}$;
we denote $|p|=\bigcup_{i}\hat\Regione_{i}$. Then the condition is
\begin{equation}
\label{aggettivo}
\bigwedge_{p}\FG(|p|)=\FG(\Regione_{0})\vee\FG(\Regione),
\end{equation}
where the intersection is taken over all paths $p$ connecting $\Regione_{0}$\
and $\Regione$.

In order to apply these results to our case, we need the following
\begin{lemma}
\label{lemma_compl} 
If $\FG_{1},\FG_{2}$ are two field nets with normal commutation 
relations, fulfilling~(\ref{aggettivo}), 
then also $\FG(\Regione)=
(\FG_{1}(\Regione)\overline\otimes\FG_{2}(\Regione))^{\text{\em Klein}}$ 
has normal commutation relations and satisfies (\ref{aggettivo}).
\end{lemma}

{\em Proof}.
Since the Klein 
transformation is normal,
\begin{eqnarray*}
\lefteqn{(\FG_{1}(\Regione_{0})\overline\otimes
\FG_{2}(\Regione_{0}))^{\klein}
\bigvee(\FG_{1}(\Regione)\overline\otimes
\FG_{2}(\Regione))^{\klein}=}\\
&=&\left[\left(\FG_{1}(\Regione_{0})\bigvee\FG_{1}(\Regione)\right)
\overline\otimes
\left(\FG_{2}(\Regione_{0})\bigvee\FG_{2}(\Regione)\right)\right]^{\klein}=\\
&=&\left[\left(\bigwedge_{p}\FG_{1}(|p|)\right)\overline\otimes
\left(\bigwedge_{q}\FG_{2}(|q|)\right)\right]^{\klein}=
\left(\bigwedge_{p,q}\FG_{1}(|p|)\overline\otimes
\FG_{2}(|q|)\right)^{\klein}=\\
&=&\bigwedge_{p}(\FG_{1}(|p|)\overline\otimes\FG_{2}(|p|))^{\klein}
\hspace{2ex}\halmos
\end{eqnarray*}

This lemma applies equally well to any finite tensor product 
$\FG^{(n)}(\Regione)=
\big(\FG(\Regione)\overline{\otimes}\dotsm\overline\otimes
\FG(\Regione)\big)^{\klein}$. 
It can be easily extended to  the infinite tensor 
product $\FG=\bigvee_n\tilde{\FG}^{(n)}$, where 
$\tilde{\FG}^{(n)}=\FG^{(n)}\otimes I\otimes I\otimes\dotsm$ acts on 
$\bigotimes_j^{\Omega_j}\mathcal H_j$, since the maps 
\[
P^{(n)}F_1\otimes F_2\otimes\dotsm=
F_1\otimes\dotsm\otimes F_n\otimes\omega_0(F_{n+1})I\otimes\omega_0(F_{n+2})
I\otimes\dotsm
\]
converge pointwise strongly 
to the identity map as $n\rightarrow\infty$, commute with the Klein 
transformations, 
and $\tilde\FG^{(n)}(\Regione)=P^{(n)}\FG(\Regione)$. 
This is precisely the case of the models constructed in section \ref{costr_F}. 
Hence, the field nets enjoy both the split property and the cohomological 
condition~(\ref{aggettivo}). Therefore the superselection structure of 
the net of local observables $\AG$ associated to the gauge triple 
$T=\{G,k,\mu\}$ is described precisely by the given group $G$.

\appendix
\section*{Appendix}
\setcounter{section}{1}
\setcounter{equation}{0}
\numberwithin{equation}{section}
We first recall the definition of the generalized free scalar field $\phi$ 
(case $D(k)=I$).

Let $J$ be a complex conjugation on $K$ commuting with $D$, and $H$ the 
subspace of real (i.e. $J$-invariant) vectors; our test functions $f$ on 
$\mathbb R^4$ take values in $H$, have range each in a finite dimensional
$D$-invariant subspace, and $(\xi,f(\cdot))$ is  
$\mathcal C^{\infty}$ with compact support for each $\xi\in K$; they carry 
actions of $G$ and $\mathcal P$ given by
\begin{eqnarray}
\label{G+}
{}_gf(x)&=&D(g)f(x),\quad g\in G,x\in\mathbb R^4,\\
\label{P+}{}_{(a,\varLambda)}f(x)&=&f(\varLambda^{-1}(x-a)),
\quad(a,\varLambda)\in\mathcal P,x\in\mathbb R^4.
\end{eqnarray}
On these test functions define the scalar product
\[
(f,g)=\sum_{\sigma}
\int (\hat f_\sigma(p),\hat g_\sigma(p))\;d\Omega^+_{\mu(\sigma)}(p),
\]
where $\hat f$ is the Fourier transform of $f$, and 
$d\Omega_m^+$ is the Lorentz
invariant measure on the positive energy mass hyperboloid such that for each 
continuous function $g$ with compact support 
\[
\int g(p)\;d\Omega_m^+(p)=
\int g(\sqrt{({\boldsymbol{p}}^2+m^2)},\boldsymbol{p})\frac{d^3\boldsymbol{p}}{2\sqrt{({\boldsymbol{p}}^2+m^2)}}.
\]
The generalized free scalar field $\phi$ is the map from our real test 
functions $f$ to selfadjoint operators $\phi(f)$ defined by the vacuum 
functional $\omega_0$ on the Weyl relations
\begin{gather*}
e^{i\phi(f)}e^{i\phi(g)}=e^{-\frac{i}{2}\text{Im}(f,g)}e^{i\phi(f+g)},\\
\omega_0(e^{i\phi(f)})=e^{-\frac{1}{4}(f,f)};
\end{gather*}
one may check
\[
\text{Im}(f,g)=\sum_{\sigma}\int\big(f_\sigma(x),
(\Delta_{\mu(\sigma)}\star g_{\sigma})(x)\big)d^4x,
\]
where, in the convolution product, $\Delta_m$, the Fourier transform of 
$\tfrac{1}{2i}(d\Omega^+_m(p)-d\Omega^+_m(-p))$,
is the usual Lorentz invariant commutator kernel with support excluding all 
spacelike points.

The generalized free Dirac field $\psi$ (case $D(k)=-I$) will be defined on 
test functions $f$ on $\mathbb R^4$ with values in $K\otimes\mathbb C^4$, each
with range in a finite dimensional $D$-invariant subspace, and such that 
$(\xi,f(\cdot))$ is $\mathcal C^{\infty}$ with compact support for each 
$\xi\in K\otimes\mathbb C^4$; they carry actions of $G$ and of
the covering group $\tilde{\mathcal P}$ of the Poincar\'e 
group\footnote{i.e. the semi direct product of $SL(2,\mathbb C)$ and the 
translations by the action $x\mapsto \varLambda(A)x$.} 
given by
\begin{eqnarray}
\label{G-}
{}_gf(x)&=&D(g)\otimes I\;f(x),\quad g\in G,x\in\mathbb R^4,\\
\label{P-}{}_{(a,A)}f(x)&=&I\otimes S(A)\;f(\varLambda(A)^{-1}(x-a)),
\quad(a,A)\in\tilde{\mathcal P},x\in\mathbb R^4,
\end{eqnarray}
where $S$ is the $(\tfrac{1}{2},\tfrac{1}{2})$ linear representation
of $SL(2,\mathbb C)$ on $\mathbb C^4$ acting on the associated Dirac 
$\gamma$-matrices\footnote{An explicit representation is e.g. given by
\[
S(A)=\begin{pmatrix}A&0\\ 0 &{A^{*}}^{-1}\end{pmatrix},\quad
\gamma_0=\begin{pmatrix}0&I\\ I& 0\end{pmatrix},\quad
\boldsymbol{\gamma}=\begin{pmatrix}0&\boldsymbol{\sigma}\\
-\boldsymbol{\sigma}& 0\end{pmatrix},
\]
where 
\[
\boldsymbol{\sigma}=
\left(
\begin{pmatrix}0&1\\1&0\end{pmatrix},
\begin{pmatrix}0&-i\\i&0\end{pmatrix},
\begin{pmatrix}1&0\\0&-1\end{pmatrix}
\right) 
\] \
are the Pauli matrices.}
by
\[
S(A)\not\!pS(A)^{-1}=\not\!p',\quad p'=\varLambda(A)p\;;\quad\quad
\not\!p:=\gamma_\mu p^\mu.
\]
On these test functions define the positive semidefinite scalar product
\begin{equation}
\label{Dir_2}
(f,g)=\sum_{\sigma}\int
\big(f_\sigma(x),i\gamma_0(-i\not\!\partial+\mu(\sigma))
\Delta_{\mu(\sigma)}\star g_\sigma)(x)\big)d^4x,
\end{equation}
and let $(f,g)_+$ be its positive frequency part, obtained by replacing in 
(\ref{Dir_2})  $\Delta_{m}$ by $\Delta^+_{m}$, the Fourier 
transform of $\tfrac{1}{2i}d\Omega_m^+$.

The Dirac generalized free field $\psi$ is defined by the GNS representation
of the pure gauge invariant quasi free state~$\omega_0$ on the CAR algebra
associated to~(\ref{Dir_2}):
\begin{gather*}
\{\psi(f)^*,\psi(g)\}=(f,g)I,\\
\omega_0(\psi(f)^*\psi(g))=(f,g)_+. 
\end{gather*}
In this way one may obtain functors $\Gamma_\pm$ whose tensor product combines,
in the general case where $D(k)$ splits $K$ into $K_+\oplus K_-$, to give the 
desired functor $\Gamma$. In particular, since the actions 
(\ref{G+},\ref{P+},\ref{G-},\ref{P-}) of $G$ and $\tilde{\mathcal P}$
on test functions induce automorphisms of the tensor product of the Weyl and 
CAR algebras on $K_+$ and $K_-$, leaving the (tensor product) vacuum state 
invariant, these actions are implemented by unitary strongly continuous
representations $\mathcal U,\mathcal V$. The von Neumann algebra 
$\mathfrak F(\mathcal O)$, $\mathcal O\in\mathcal K$, is generated by 
$\exp{i\phi(f)}\otimes I$ and $I\otimes \psi(g)$, $f,g$ varying over the test 
functions associated respectively to $K_+$ and $K_-$ as above, with support in
$\mathcal O$.

The case of generic spins and possibly continuous mass spectrum
, not used 
in this paper, might be handled in a similar way.

\end{document}